\documentclass[11pt,twoside]{article}
\usepackage{asp2010}

\aspvolume{495}
\aspcpryear{2015}
\aspvoltitle{Astronomical Data Analysis Software and Systems: XXIV}
\aspvolauthor{A. R. Taylor and E. Rosolowsky, eds.}

\resetcounters

\begin{document}

\title{Data Analysis Software for the ESPRESSO Science Machine}
\author{Guido~Cupani$^1$, Valentina~D'Odorico$^1$, Stefano~Cristiani$^1$, Jonay~Gonz\'alez-Hern\'andez$^2$, Christophe~Lovis$^3$, S\'ergio~Sousa$^4$, Eros~Vanzella$^5$, Paolo~Di~Marcantonio$^1$, and Denis~M\'egevand$^3$
\affil{$^1$INAF--OATs, Via Tiepolo 11, 34143 Trieste, Italy}
\affil{$^2$IAC, V\'ia L\'actea, 38205 La Laguna, Tenerife, Spain}
\affil{$^3$Universit\'e de Gen\`eve, 51 Chemin des Maillettes, 1290 Versoix, Switzerland}
\affil{$^4$CAUP, Rua das Estrelas, 4150-762 Porto, Portugal}
\affil{$^5$INAF--OABo, Via Ranzani 1, 40127 Bologna, Italy}}

\begin{abstract}
ESPRESSO is an extremely stable high-resolution spectrograph which is currently being developed for the ESO VLT. With its groundbreaking characteristics it is aimed to be a ``science machine'', i.e. a fully-integrated instrument to directly extract science information from the observations. In particular, ESPRESSO will be the first ESO instrument to be equipped with a dedicated tool for the analysis of data, the Data Analysis Software (DAS), consisting in a number of recipes to analyze both stellar and quasar spectra. Through the new ESO Reflex GUI, the DAS (which will implement new algorithms to analyze quasar spectra) is aimed to get over the shortcomings of the existing software providing multiple iteration modes and full interactivity with the data.\end{abstract}

\section{Introduction: a science machine for the VLT}
ESPRESSO (Echelle SPectrograph for Rocky Exoplanets and Stable Spectral Observations, \citeauthor{2014AN....335....8P}, \citeyear{2014AN....335....8P}) is a fiber-fed, cross-dispersed echelle spectrograph to be commissioned in the Combined-Coud\'e Laboratory (CCL) of the ESO VLT in the Paranal Observatory in Chile, starting from 2016. It is designed to achieve exceptional standards of precision, resolution, and stability, which are motivated by two driving science cases: (i) the search for rocky exoplanets in the habitable zone around stars with the radial velocity technique \citep{2003Msngr.114...20M}, and (ii) the study of the variability of fundamental constants through the observations of the absorption systems along the line of sight to distant quasars (QSOs). These requirements put strong constraints to the instrument design. A two-arm layout was chosen, with two large science detectors ($\sim$$90$ mm $\times$ $90$ mm) covering a spectral range from $380$ to $780$ nm in the visible band. The optical bench, free of movable parts, will be enclosed in a vacuum vessel and insulated by two thermal chambers, to reach a pressure stability of $\sim$$5$ $\mu$bar and a temperature stability of $\sim$$1$ mK at the echelle grating. A laser frequency comb will be used for wavelength calibration, to achieve a radial velocity precision of $10$ cm s$^{-1}$. Finally, exploiting its location, the instrument has been designed to work with any of the VLT unit telescopes (UTs) available, in high-resolution mode ($R\sim 134,000$) or ultrahigh-resolution mode ($R\sim 225,000$); or with all four UTs at the same time, in medium-resolution mode ($R\sim 59,000$). Given these characteristics, ESPRESSO is expected to have a significant impact in several other areas of astronomical research, including but not limited to the study of the physical and chemical state of the intergalactic medium (IGM) at $z>2$ (see Sect.~\ref{P2-8_qso}). 

Since the beginning, ESPRESSO has been developed as an end-to-end ``science machine'', i.e. an instrument capable of providing the astronomers with high level science products within minutes after the end of the observations. To this aim, the ESO Data Flow System (DFS) already implemented at Paranal will be integrated with an ensemble of new software packages, namely: (i) the ESPRESSO Observation Preparation Software (EOPS), which will help the visitor astronomers to optimize their observations on the fly; (ii) the Data Reduction Software \citep[DRS,][]{P2.3_adassxxiii}, responsible for removing the instrumental signature from the observations; and (iii) the Data Analysis Software (DAS), responsible for extracting scientific information from the reduced data. This paper focuses on the DAS, whose first public release is expected for May, 2015. The novel approach behind the software is described, as well as some details of the analysis of QSO spectra. 

\section{Designing the ESPRESSO DAS}
The very specific science cases of ESPRESSO, combined with the requirements of the ``science machine'' concept, strongly called for a dedicated tool for data analysis. So far, ESO instruments have been equipped only with data reduction software in the form of pipelines, each one consisting in a cascade of tasks (``recipes''). This sequential approach is not well suited to data analysis, which is usually performed through several iterations of a variable number of tasks. Six tasks were identified for stellar spectra: (i) measure of the radial velocity with the cross-correlation method; (ii) computation of the stellar activity indexes; (iii) interpolation of the stellar continuum; (iv) comparison of the observed spectra with rotationally-broadened synthetic spectra; (v) measure of the equivalent width of absorption lines; (vi) estimation of the effective temperature and [Fe/H] metallicity (only for FGK stars). Three tasks were identified for QSO spectra: (i) determination of the continuum emission; (ii) Voigt profile fitting of the absorption lines; (iii) identification of the absorption systems. To guarantee a complete automation of the latter tasks while maintaining full control over the process, a great degree of interaction and flexibility must be allowed; the user should be able to inspect the results at any time, to tune the parameters, and to iterate the tasks freely; also, the partial products of any task should be fed to the other ones, to improve the results (see Sect.~\ref{P2-8_qso}).

\articlefigure{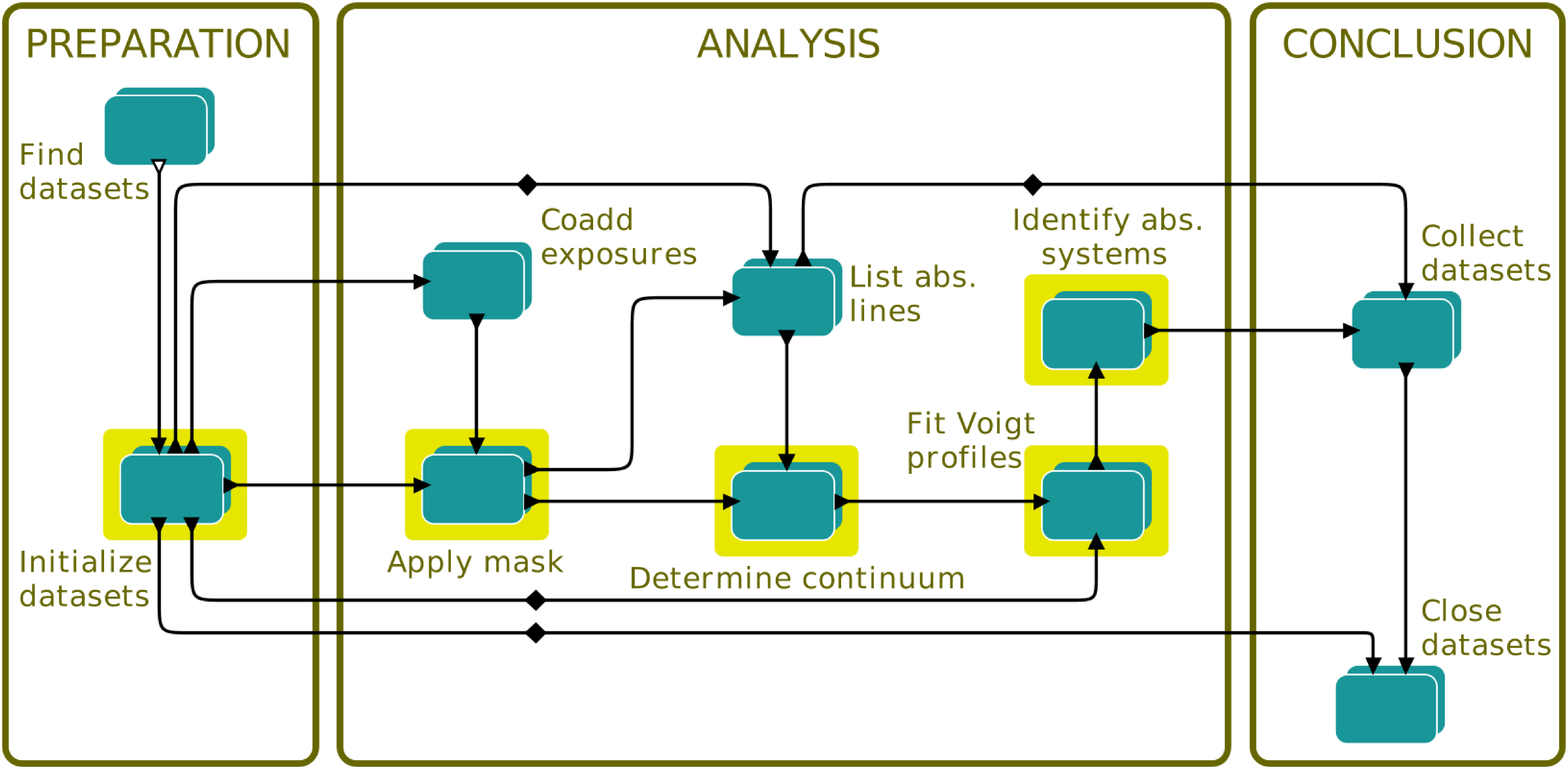}{P2-8_reflex}{ESO Reflex workflow for the QSO branch of the DAS.}
The ESO Reflex environment \citep{2013A&A...559A..96F} was adopted as a way to mediate between the standard pipeline formalism and the practical needs of the ESPRESSO DAS users. Reflex is primarily a GUI to the ESO pipelines (traditionally run from the command line) which takes care of both the organization of the I/O data and the execution of the recipes. In Reflex, a pipeline is represented graphically as a ``workflow'', provided with built-in tools to interact with the data. Through the workflow metaphor, Reflex can handle multiple iteration schemes not easily implemented from the command line. The ESPRESSO DAS is the only ESO pipeline so far to integrate Reflex features as fundamental design requirements. The pipeline itself has a multi-layered structure whose core is the ESPRESSO Data Analysis Library (DAL), a set of modules written in ANSI-C to manage basic analysis tasks (e.g. re-binning of spectra, masking of spectral regions, spectral line detection, etc.). DAL functions make extensive use of the ESO Common Pipeline Library \citep[CPL;][]{2004SPIE.5493..444M} and the GNU Scientific Library \citep[GSL;][]{galassi_2009} and constitute the ``building blocks'' of the recipes operated by Reflex. Alongside, Reflex also operates Python modules to visualize the data and set up parameters. All the low-level modules are designed to be consistent with the workflow approach and assume Reflex as the preferred interface.

Two different Reflex workflows will be developed, one for the star branch and one for the QSO branch. A total of seven out of twelve recipes have been coded as of October, 2014, including recipes to perform tasks common to both branches (co-addition of different exposures; application of a spectral mask; creation of a list of absorption lines). Fig.~\ref{P2-8_reflex} shows the first implementation of the QSO branch workflow. Depending on the input data, the correct path along the arrows is selected in the ``preparation'' phase, and one or more steps are executed. Boxes in the ``analysis'' phase correspond to pipeline recipes; highlighted boxes contain interactive Python modules. In the ``conclusion'' phase, products are collected and saved according to the requests of the user. 

\section{Data Analysis of quasar spectra}
\label{P2-8_qso}
Despite its instrument-oriented nature, the QSO branch of the ESPRESSO DAS aims to put forward a new approach in quasar spectra analysis, which is naturally generalized to all high-resolution data in the visible band (e.g. VLT UVES, Keck HIRES). The main issue to be faced when studying the IGM with a QSO as a background source is to disentangle the absorption features of the structures along the line of sight from the intrinsic emission of the QSO itself. This problems involves the interpolation of at least three different components (the non-thermal emission of the AGN; the broad emission lines of the accretion disk; the absorption lines of the IGM) and is hardly solved in a step-by-step way. In many cases, continuum is still interpolated by eye (a time-consuming method affected by large subjective bias). 

The algorithm designed for the ESPRESSO DAS recipe \texttt{espda\_fit\_qsocont}, on the contrary, is fully automatic and model-independent. The basic goal of the algorithm is to interpolate both the QSO continuum and the absorption lines at the same time. In the Lyman-$\alpha$ forest, a guess continuum is estimated enhancing the observed flux by an effective optical depth term $\tau_\mathrm{eff}(z)$; this guess continuum is iteratively refined by gradually fitting the absorption lines and taking the contribution of the fitted lines out of $\tau_\mathrm{eff}$. A test version of the algorithm has been applied to simulated and observed data at high and medium resolution (Fig.~\ref{P2-8_cont}) with promising results up to redshift $z\sim 4$, where the Lyman-$\alpha$ forest show significant line blending. The results are virtually bias-free, as they do not depend on theoretical modeling (except for the optical depth term, which becomes gradually negligible as the iteration proceeds).

The described algorithm relies on a dedicated module to fit absorption lines, which runs also as an independent recipe \texttt{espda\_fit\_voigt}  (already coded). Lines are fitted by minimizing the reduced $\chi^2$ between the observed profile and a Voigt profile, depending on line redshift, column density, and Doppler parameter. The same procedure is employed by other software packages like FITLYMAN for ESO-MIDAS \citep{1995Msngr..80...37F} and VPFIT (\copyright\ 2014 R.~F.~Carswell); compared to these tools, the ESPRESSO DAS is meant to be faster and more user friendly, thanks to the Reflex GUI. A dedicated recipe \texttt{espda\_iden\_syst} to identify systems of associated absorption lines will be also provided, adapting the algorithm by \citet{1975ApJ...198...13A}.
\articlefigure{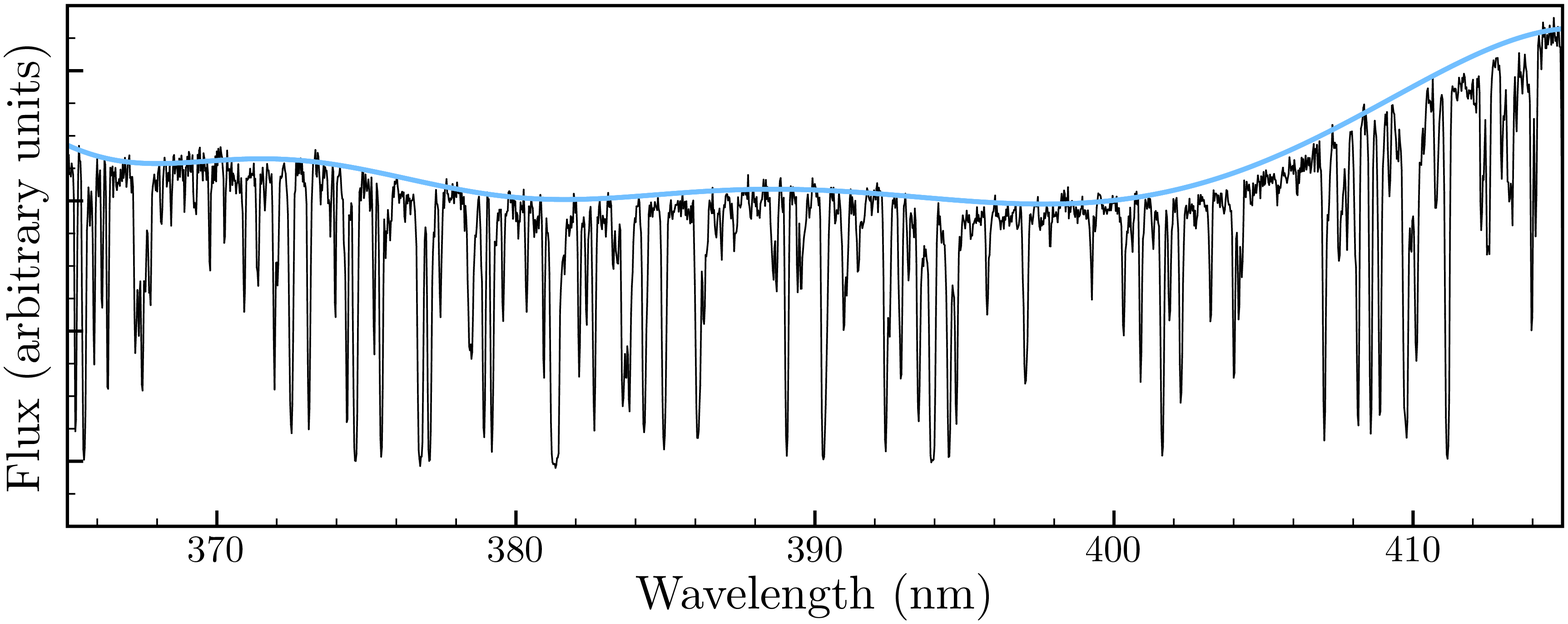}{P2-8_cont}{Determination of the continuum component (bold line) in the Lyman-$\alpha$ forest of QSO SDSS J011150.07+140141.3 (narrow line). VLT X-shooter data.}

\bibliography{P2-8}

\end{document}